\documentclass{acm_proc_article-sp}

\usepackage{url}
\usepackage{epsfig,endnotes,caption}
\usepackage{amssymb,amsmath, stmaryrd}
\usepackage[lofdepth,lotdepth]{subfig}

\makeatletter
\def\@copyrightspace{\relax}
\makeatother

\begin{document}


\title{Albatross: a Privacy-Preserving Location Sharing System}
\subtitle{Extended version of ASIACCS 2015 paper}

\numberofauthors{3} 
\author{
%
%
\alignauthor
Gokay Saldamli\\
       \affaddr{Samsung Research America}\\
       \affaddr{665 Clyde Avenue}\\
       \affaddr{Mountain View, CA 94043}\\
       \email{gokay.s@samsung.com}
\alignauthor
Richard Chow\\
       \affaddr{Intel Corporation}\\
       \affaddr{3600 Juliette Lane}\\
       \affaddr{Santa Clara, CA 95054}\\
       \email{richard.chow@intel.com}
\alignauthor
Hongxia Jin\\
       \affaddr{Samsung Research America}\\
       \affaddr{665 Clyde Avenue}\\
       \affaddr{Mountain View, CA 94043}\\
       \email{hongxia.jin@samsung.com}
}

\maketitle

\begin{abstract}
Social networking services are increasingly accessed through mobile devices. This trend has prompted services such as Facebook and Google+ to incorporate location as a de facto feature of user interaction. At the same time, services based on location such as Foursquare and Shopkick are also growing as smartphone market penetration increases. In fact, this growth is happening despite concerns (growing at a similar pace) about security and third-party use of private location information (e.g., for advertising). Nevertheless, service providers have been unwilling to build truly private systems in which they do not have access to location information. In this paper, we describe an architecture and a trial implementation of a privacy-preserving location sharing system called Albatross. The system protects location information from the service provider and yet enables fine-grained location-sharing. One main feature of the system is to protect an individual's social network structure. The pattern of location sharing preferences towards contacts can reveal this structure without any knowledge of the locations themselves. Albatross protects locations sharing preferences through protocol unification and masking. Albatross has been implemented as a standalone solution, but the technology can also be integrated into location-based services to enhance privacy.
\end{abstract}


%


%

\section{Introduction} \label{intro} 

Smartphones are likely to become the fastest-spreading technology in history, beating even electricity and television. According to a March 2012 PEW report~\cite{S12:46}, nearly half (46\%) of American adults were smartphone owners and by 2017 three billion of the estimated nine billion worldwide mobile subscriptions will use a smartphone. With the spread of smartphones, applications with location capabilities will increase in popularity. Location-based services (LBS) typically offer benefits such as precise navigation, location-based discount coupons, or easy information sharing through features like social check-ins. These services are likely to reach hundreds of millions of users over the next couple of years with the help of smartphone market penetration.

In this paper, we concentrate on a particular LBS, social location sharing. With services such as Foursquare~\cite{Foursquare}, users can checkin and share their current location with contacts. Status updates which include location has now become standard on general social networks such as Facebook and Google+. Applications such as Google Latitude~\cite{Latitude}, Highlight~\cite{Highlight}, Apple's Find My Friends~\cite{FindMyFriends}, and (reportedly) Facebook~\cite{FacebookLocation} go a step further by sharing location with friends even while running in the background.

According to another PEW report~\cite{PEW_Location}, 18\% of all smartphone owners and 10\% of all adults share location. Foursquare claims to be used by nearly 30 million people worldwide~\cite{Foursquare}. Such growth has not come without concern; in many ways, location sharing has been a poster child for online privacy. For instance, the site PleaseRobMe.com was launched to list people who were not home according to their Foursquare checkins~\cite{Foursquare}. According to~\cite{PEW_Privacy}, over 30\% of smartphone users have turned off the location tracking feature on their cell phone because they were concerned that other individuals or companies could access that information. 

The work described in this paper attempts to address the privacy concerns of location sharing. We describe the architecture of Albatross, a location sharing application designed for privacy. The system is named after one of the most mobile of animals: ``Albatrosses range over huge areas of ocean and regularly circle the globe.''~\cite{albatross}.

One goal of Albatross is to hide the user's location from the server and yet allow flexible sharing with other users. Albatross can be used for sharing users' whereabouts with only selected contacts, such as close friends, family members and colleagues. At the same time, another goal of Albatross is to hide the user's privacy policies for his location. These policies may be even more sensitive than the location itself, as the policies (i.e., user's sharing preferences) can imply who is favored by the user and who is not. We hide or mask a user's sharing preferences by using a meta-protocol that hides our individual protocols. We also obscure traffic activity through dummy sharing events.

Albatross supports an asynchronous sharing model, where Alice can share her location or checkin to an offline contact who can then see Alice's location after coming online. Naturally, Albatross can also be used for automated periodic sharing or ``tracking,'' say sharing location with a spouse.

As Albatross never shares the location information with the server, it is markedly different from current location sharing services such as Foursquare and Google Latitude. All of our protocols protect users' location from the server, and in addition regulate and conceal the patterns of sharing among users.





\subsection{Contribution}
Albatross is a location-sharing system composed of several simple cryptographic protocols that allow users to share their locations with their contacts privately. Albatross keeps key features of a typical location-sharing service while promoting privacy. The following are the two main goals:
\begin{itemize}
\item {\bf Location privacy.} Users share their location information only with their contacts; location data is always encrypted and never visible to the service provider. 
\item {\bf Social Network Privacy.} By virtue of routing all traffic, a service provider can typically observe a user's active social network and temporal habits. For example, which contacts a user shares most often with and which contacts a user shares with at different times of day. Albatross protects against traffic analysis by protocol hiding and adding dummy traffic. 
\end{itemize}


Albatross is practical in terms of performance. All of our protocols are lightweight, using symmetric key encryption, and can easily be run on a mobile phone. Nevertheless, Albatross does introduce considerable overhead compared to a non-private, trusted server system. Hence, we have emphasized efficiency in our design and made choices that reduce computation and storage when possible. One example is the introduction of a novel grid scheme in Section~\ref{sec:grid} that requires only one protocol run compared to multiple runs in other schemes, albeit at a small reduction in privacy. We implemented Albatross as an Android application for smart phones and measured performance overhead of the system.

\begin{figure}[htbp]
\centering
\subfloat[][Checkin screenshot]
{
    \includegraphics[trim = 0mm 0mm 102mm 0mm, clip, scale=0.35]{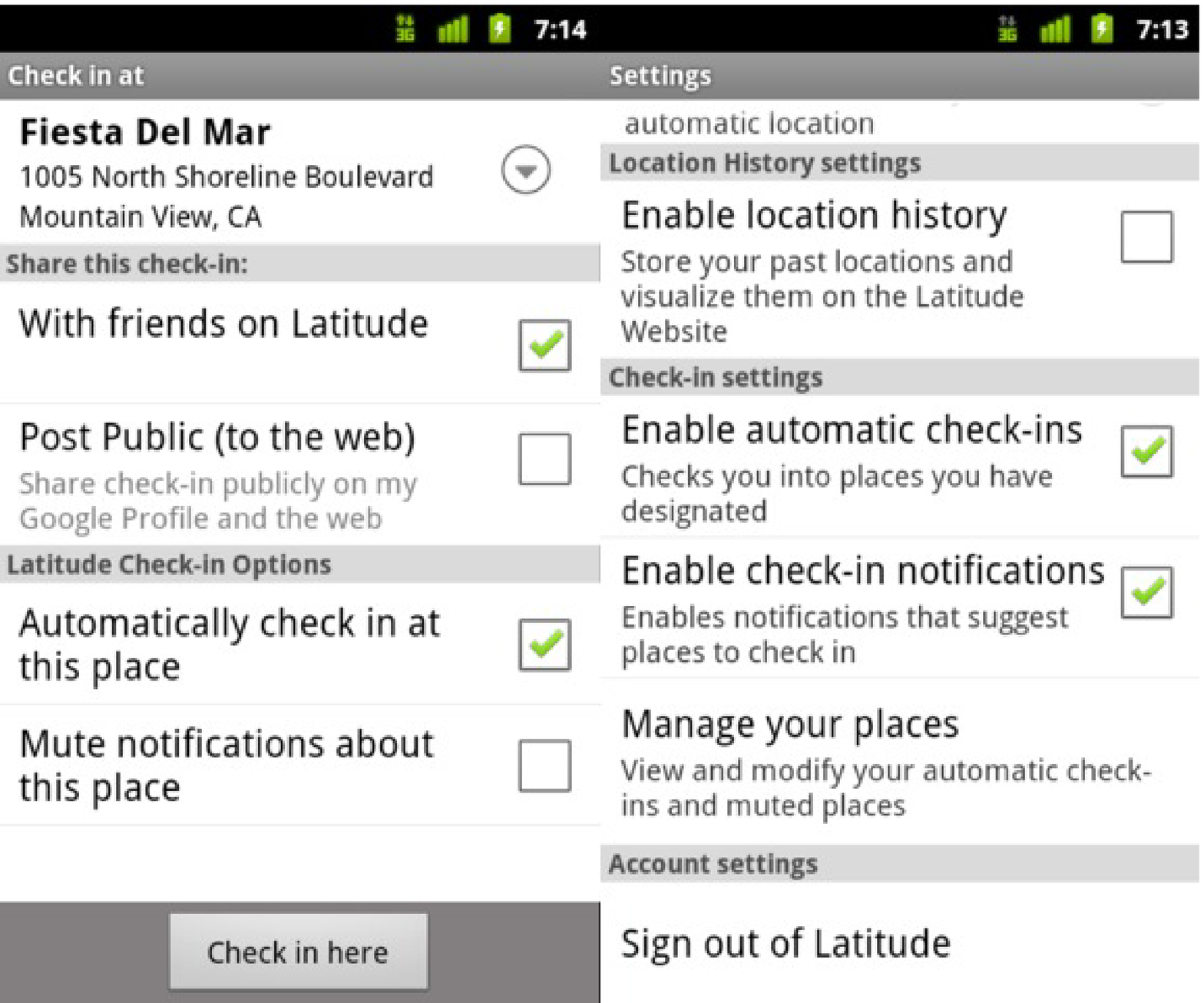}
    \label{fig:lat-checkin}
}
\qquad
\subfloat[][Location reporting]
{
    \includegraphics[trim = 0mm 30mm 0mm 0mm, clip, scale=0.35]{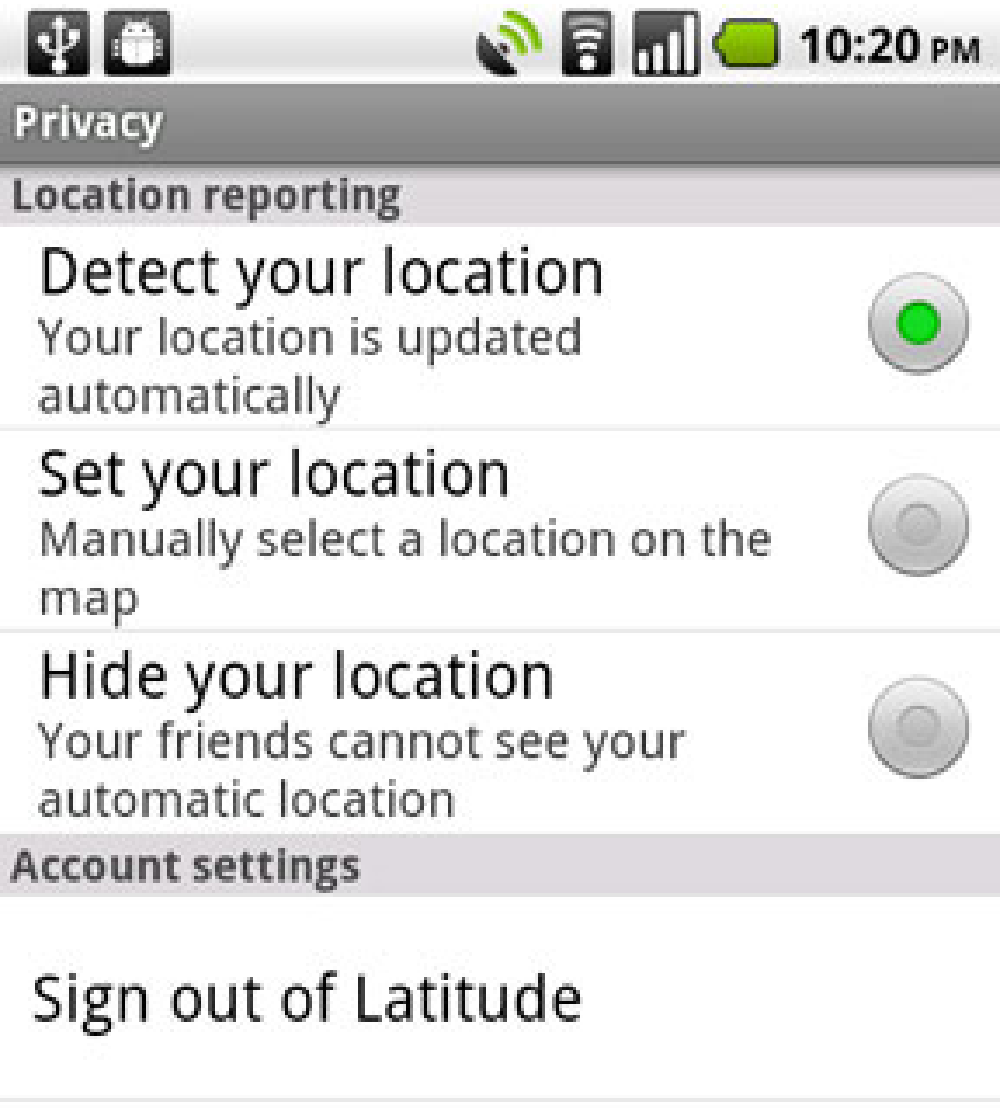}
    \label{fig:lat-sharing}
}
\caption{Location sharing with Google Latitude.}
\label{fig:lat-checkinandsharing}
\end{figure}

The rest of this paper is organized as follows. Section~\ref{sec:SharingService} is a brief introduction to location-sharing services. Section~\ref{sec:privacygoals} describes our privacy model and goals. Section~\ref{sec:protocols} overviews the base location-sharing and cryptographic building blocks used to construct Albatross. Section~\ref{sec:grid} describes our grid and cell system. Section~\ref{sec:unified} details the main Albatross protocol, one that incorporates the basic protocols of Section~\ref{sec:protocols}. Section~\ref{sec:sys} describes our prototype implementation.

\section{Location Sharing Services}
\label{sec:SharingService}
We provide some background on location sharing services.

\subsection{Example of Google Latitude}
In general, users run a mobile application that connects to the location sharing service website. Through the application, the user can {\it checkin} or announce his location to some set of his contacts. Two users can become contacts by mutual agreement; for instance, one user can invite another user to be a contact, which can then be accepted. We briefly outline the functionality of Google Latitude~\cite{Latitude}, which is fairly typical of location sharing services. Figure~\ref{fig:lat-checkin} shows a screenshot of checkin. Note that in Google Latitude a user can checkin to a fake location (see Figure~\ref{fig:lat-sharing}).

\begin{figure}[htbp]
  \centering
\subfloat[][Sharing preference for a particular friend]
{
    \includegraphics[scale=0.35]{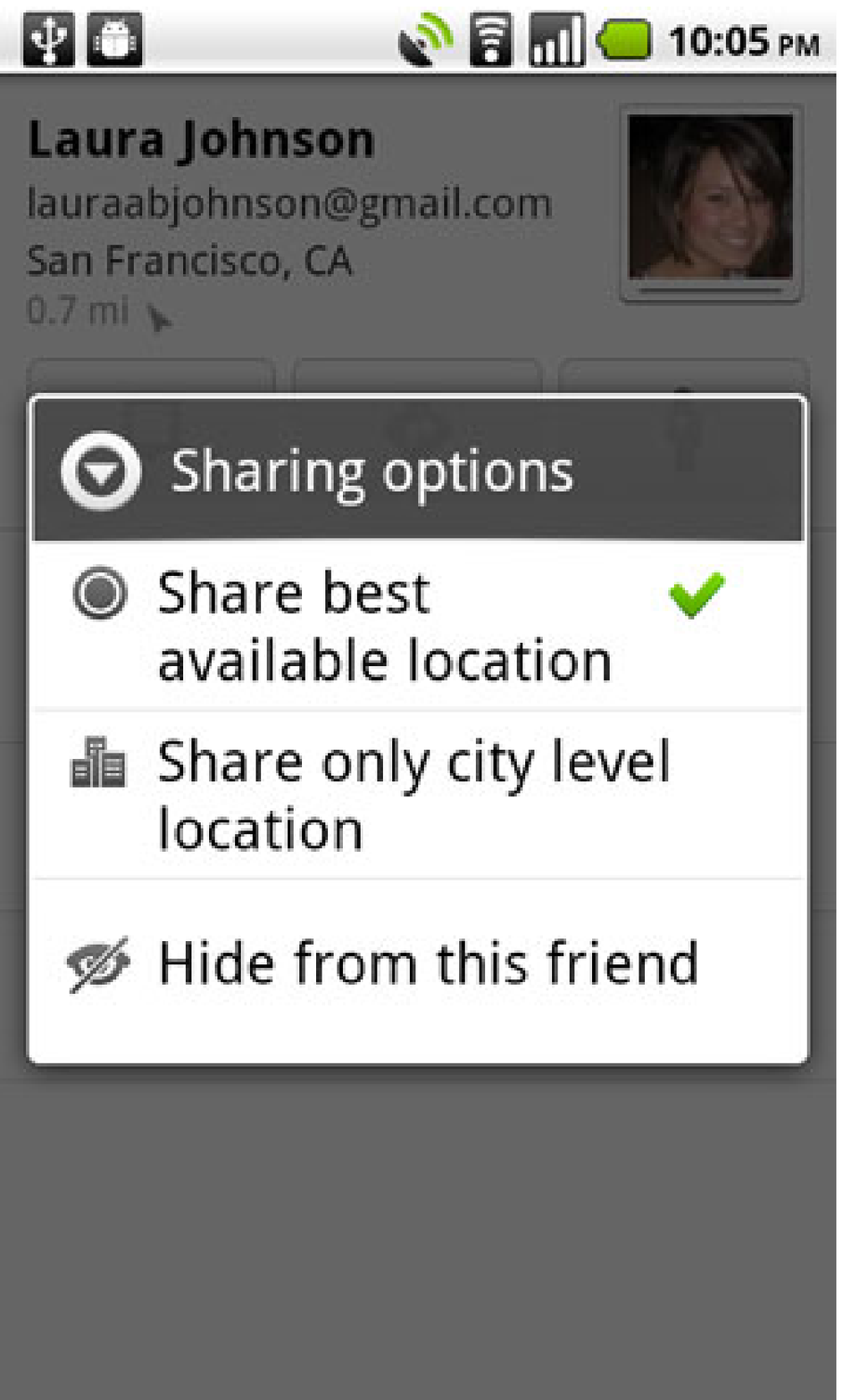}
    \label{fig:lat-control}
}
\qquad
\subfloat[][Retrieval of friend locations]
{
    \includegraphics[scale=0.35]{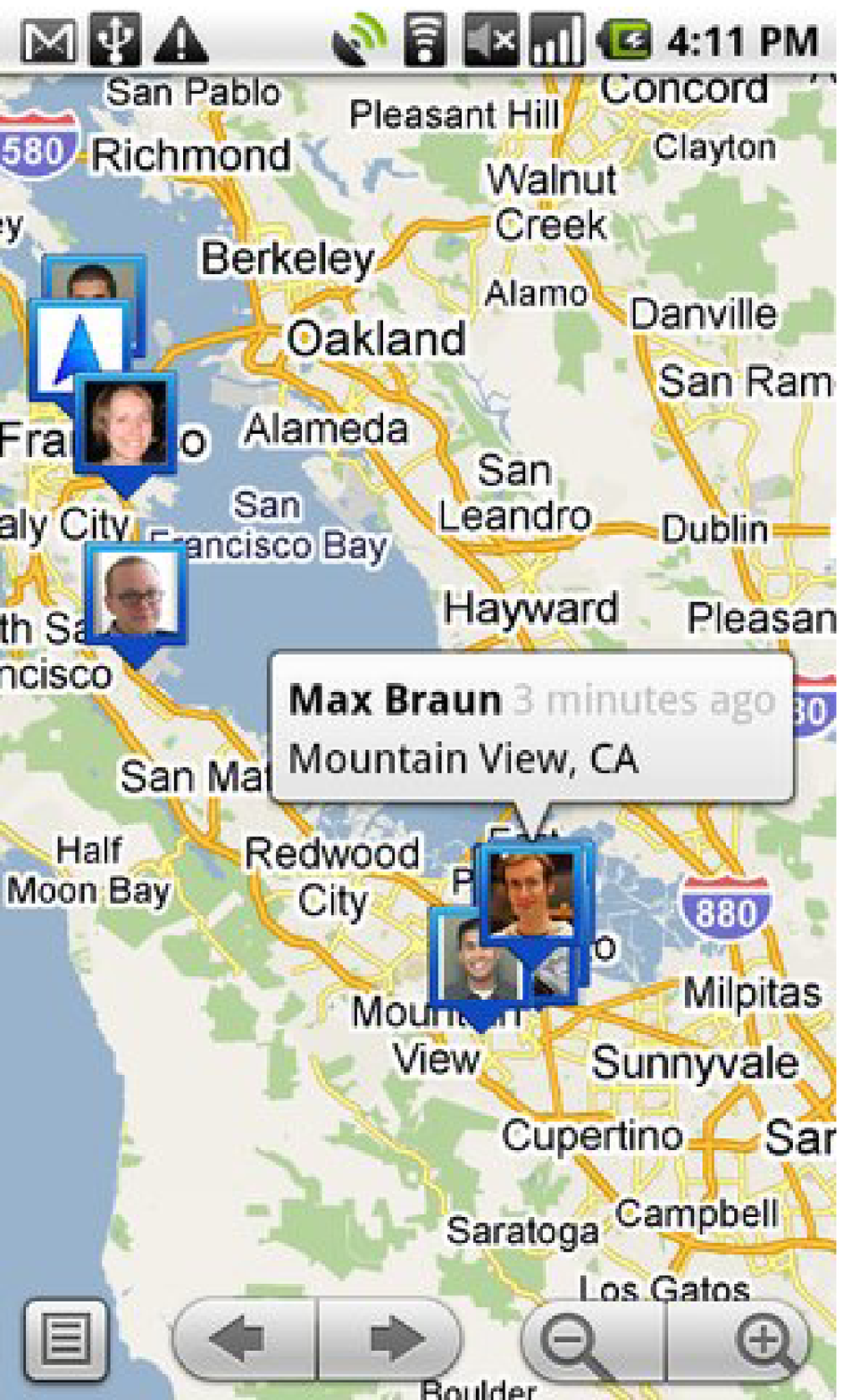}
    \label{fig:lat-see}
}
\caption{Friends in Google Latitude}
\label{fig:lat-controlandsee}
\end{figure}

For each contact, a user may set a different preference or {\it granularity} of location sharing. See Figure~\ref{fig:lat-control} for the example of Google Latitude. A user can share his exact location, an approximate location, or not share location at all. Finally, using an interface such as Figure~\ref{fig:lat-see}, a user can {\it retrieve} or see the locations of his friends.

Note that the locations of users are visible to Google Latitude. Furthermore, the sharing activity and granularities for each user are also available. Hence, Google Latitude can reconstruct not only users' movements over time but the whole social network.


\subsection{Location Sharing Granularities}
Location sharing preferences can be quite diverse; for instance, users may want to share their exact location, an approximate location (e.g., city-level as in Figure~\ref{fig:lat-control} in Google Latitude), choose to be invisible (``Hide our location'' in Figure~\ref{fig:lat-sharing}), or provide a fake location (``Set your location'' in Figure~\ref{fig:lat-sharing}). Each choice is called the {\it granularity} of the location-sharing. The granularity of sharing may vary with the particular contact. For instance, a user may want to share their exact location with family, an approximate location with close friends, and be invisible to everyone else. These preferences may also change from moment to moment.

Albatross offers a set of granularities \{available, approximate, nearby, invisible, fake\} composed of the most common preferences. We summarize the meaning of these granularities as follows:
\begin{itemize}
\item Available: share exact location
\item Approximate: share approximate location, e.g., the city or region
\item Nearby: share approximate location only if the contact is nearby
\item Invisible: share nothing, also known as {\it hiding}
\item Fake: share a fake location. This might be an exact or approximate location.
\end{itemize}

\section{Location-Sharing Privacy Definition}
\label{sec:privacygoals}
A location-sharing service manages events of the following form: User A shares timestamped location information (of some granularity) with User B. We assume the service relays traffic between User A and User B and hence knows the time of the sharing event. Historical data for the service consists of a record of all events with timestamps.

Sharing events may be encrypted so that the service cannot necessarily determine the location or granularity of the sharing event but does know the parties involved in the sharing event. We also assume the possibility of ``dummy'' sharing events that do not convey any information from User A to User B. The service cannot distinguish between an actual sharing event and a ``dummy'' event. Hence, despite the service observing all the traffic, the service does not know whether a user has actually shared location information with another user. 

Assuming the service has all historical data to analyze, ideal privacy for users of a location-sharing service consists of the following:

\begin{itemize}
\item At any time, the service knows nothing about any user's location.

\item At any time, the service or any other users know nothing about whether any user is online. This implies a lack of gaps in the event stream for any user. Such a gap may imply undesirable inferences, such as an airplane flight.

\item At any time, only users which User A has shared with know about A's location. These users know User A's location only to the degree of granularity specified.

\item The service learns nothing about a user's contacts. In particular, the service learns nothing about a user's social circles, e.g., who he commonly shares with and to what degree.  
\end{itemize}

In practice, ideal privacy is difficult to achieve since in our architecture all traffic goes through the server. A user's active contacts would eventually be revealed through traffic analysis, unless a large amount of dummy traffic is sent. In our design we chose to reveal each user's contacts to the server. This seems an acceptable tradeoff for scalability, as securing location is the primary goal and the inner structure of each user's contacts (e.g., close vs.\ not-so-close friends) remains protected.

Likewise, we chose not to protect temporal patterns from the server. This is, however, an interesting direction for future research. A system where every device appears active to the server is possible; one scheme would involve an auxiliary proxy (for instance, in the cloud or a desktop at home) that funnels traffic and covers up for the device if it is offline.

\subsection{Adversarial Model}
Before we discuss the privacy goals of Albatross, we outline the adversarial model. We assume an untrusted server, but a server that is Honest-but-Curious. This assumption is realistic, as argued in~\cite{Hummingbird}: violating the honest-but-curious model risks damage to a service provider's reputation. The server correctly routes protocol messages, but does not attempt to create fake users or collaborate with existing users. In particular, for our VPET protocol (see Section~\ref{sec:vpet}), server collusion with User B would reveal User A's location.

We see data sub-poenas, hacker break-ins, or inadvertent data release as examples of threats on server-side data. These are the main motivation for the Albatross system. Threats from users trying to infer the location of another user also must be considered, although for the most part these threats are similar to threats in existing systems. In the Albatross system, the attacker would have to be a contact of another user to get any information at all. However, for the ``nearby'' location granularity (in which a user reveals location to nearby contacts only), there are new privilege escalation attacks, where an attacker granted the nearby location privilege tries virtually placing himself in many locations in order to narrow down the possible location space of the user.

\subsection{Privacy Goals of Albatross}
\label{sec:AlbatrossPrivacyGoals}
Our privacy goals improve upon existing systems in that the server remains oblivious to users' (1) locations, and (2) sharing granularities or circles. Specifically, our goals are:
\begin{itemize}
\item {\bf P1 (Server):} The server learns nothing other than the set of contacts for each user and which users are online in any given time interval. The server does not learn anything about a user's social network structure (other than the set of contacts). The server does not learn the granularity of shared location. 
\item {\bf P2 (User):} A user's contacts learn only the location granularity that he intends for them to know. This goal is no different from current location sharing services such as Google Latitude. Furthermore, his contacts cannot infer when a user is offline. The reason is that when User A is invisible to User B, User A wants there to be uncertainty as to whether  A is being purposefully invisible or simply offline.
\end{itemize}

Note that network and device information may reveal information about a user; for instance, the IP address may reveal an approximate location using a geo-location database. Similarly, a mobile carrier will have location information on a device from the cell towers it connects to. These concerns may be mitigated through use of an anonymizing network and a proxy device, but we do not directly address these concerns in this paper and assume the server's information consists solely of our protocol information.

\section{Base Protocols}
\label{sec:protocols}

We start with the case where User B shares his exact location with User A at a certain time; in other words, User B is checking in to User A with the preference of ``available.'' User A then retrieves User B's location information, perhaps at some later point in time. We hide this sharing from other contacts and the server with the following simple asynchronous protocol. The same protocol is used if B shares an approximate or a fake location. 

We assume the existence of a shared key $k_{AB}$ between users who are contacts $A$ and $B$. This shared key can be created at the time a user adds another user as a contact. If a public/private key pairs is in place on the device, say through an existing PKI set up by the social network, this shared key can be formed using the Diffie-Hellman key-exchange protocol. SocialKeys has been proposed as a lighter weight alternative~\cite{NTL11:Location}.

\subsection{Private Sharing Protocol (PSP)}
With PSP, the location information -- either approximate, exact, or fake -- is simply encrypted so that it remains hidden from unauthorized users and the server. Let A-Alice and B-Bob and S-Server be the three parties involved in the protocol. Table \ref{tbl:provalues} gives the basic parameters used in PSP.

\begin{table} [htbp]
\renewcommand\arraystretch{1.2}
\small
\centering
\caption{Basic protocol parameters} \label{tbl:provalues}
\begin{tabular}{|c|c|} \hline
$x_A$ and $x_B$ & Location information for A and B respectively \\ \hline
$E$ & Symmetric encryption function \\ \hline
$k_{AB}$ & is the key shared by users $A$ and $B$\\ \hline
 & a counter value set to zero in the very first \\ 
$ctr$ & handshake between $A$ and $B$ and incremented \\ 
& whenever new masking keys are needed. \\ \hline
$k_1, k_2$\ldots$$ & parts of the encrypted $ctr$ value \\ \hline
\end{tabular}
\end{table}

\begin{figure}[htbp]
  \centering
    \includegraphics[scale=0.45]{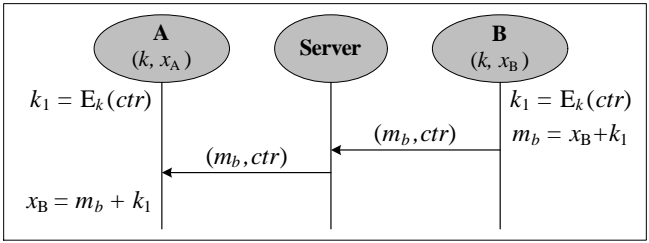}
    \caption{Private sharing protocol}
    \label{fig:PSP}
\end{figure}

Figure~\ref{fig:PSP} shows how the location information, $x_B$ is encrypted by user $B$ so that it remains hidden from the server. The encryption is performed by masking the location with the output of the pseudorandom function (PRF) $E$. Since the bit size of the mask is determined by the location precision, a single encryption can be used for multiple sharing instances. For instance, if the location information $x_B$ is presented by 32-bits and $E$ is 128-bit block cipher such as AES, both parties would perform a single encryption of $ctr$ and reuse the four different blocks of the ciphertext $k_i$. This is equivalent to a single encryption after every four sharing instances. Note that the Server is passive, i.e., it only transmits the messages in this protocol.

In some cases, users might want to share their location information with only their nearby contacts (e.g., within a circle of radius r) who have made their location visible (available, approximate, or nearby). Compared to unconditional sharing in PSP, this sharing has the requirement that users need to know their contact's location or proximity before sharing their own location. For instance, if user $A$ receives location information $x_B$ from user $B$, he may determine that user $B$ is nearby and decide to share his location information $x_A$. In fact, this is a natural interaction in a location sharing service. Hence, PSP can still be used after checking a contact's location or proximity.


However, proximity sharing is a bit more complicated if two users want to share their locations to each other {\it only} if the other is nearby. In this case, since both parties do not have any prior location or proximity information PSP is not useful. There are more sophisticated solutions available:
 
\subsection{Vectorial Private Equality Testing (VPET) Protocol}
\label{sec:vpet}

We describe in Section~\ref{sec:petgrids} the relationship between private proximity sharing and the so-called Private Equality Testing problem (PET). Basically, one can subdivide the earth's surface into a grid and test equality of belonging to the same grid. PET is also known as "Socialist Millionaire Problem" \cite{JY96:Proving} where two millionaires want to determine whether their wealth is equivalent or not, without disclosing any information about their riches to each other. 


PET is a well-studied problem in the literature. Various aspects and use cases of PET were considered in a number of studies including \cite{FNW96:Comparing,AD01:Secure,ZGH07:Louis,NTL11:Location}. The majority of these solutions are two-party protocols using computationally intensive cryptographic primitives such as homomorphic, commutative, or public-key encryption. 
Nevertheless, there are three-party protocols such as~\cite{NTL11:Location,SCJ13:Private} that use comparably lightweight symmetric-key primitives requiring far less communication.
In Albatross we use VPET~\cite{SCJ13:Private}, one of these recently proposed protocols for sharing locations to only nearby users. Naturally, using a different PET protocol would not effect the overall system. For completeness, we briefly describe the VPET protocol here. 

\begin{figure}[htbp]
  \centering
    \includegraphics[scale=0.4]{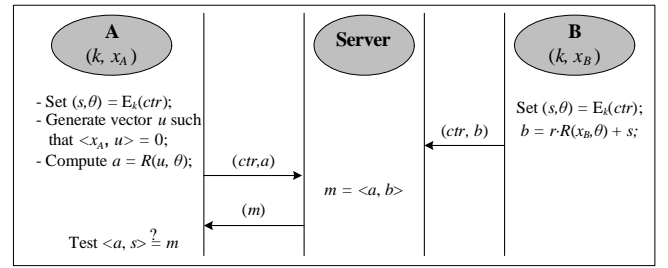}
    \caption{VPET protocol}
    \label{fig:VPET}
\end{figure}
The flow of the protocols presented in Fig. \ref{fig:VPET}. Assume that Bob wants to share his private location $x_B$  with Alice. In VPET, location values are mapped to vectors using a vectorization process and these vectors are masked by the following values
\[(s, \theta) = E_{k_{AB}}(ctr)\]
Bob arranges the output of the PRF, $E$, in a way that $s$ represents a vector having random entries and $\theta$ represents a rotation angle. 
Next, Bob computes: 
\[b = r R(x_B, \theta) + s\]	
where $r$ is a non-zero random number and $s$ is taken from the $E$ function. He sends $b$ and $ctr$ to the server.

Alice, in order to compute whether Bob is nearby, first queries the server to obtain the latest value of $ctr$ from Bob. Alice aborts if the $ctr$ received from the server is not fresh, otherwise Alice computes $(s, \theta) = E_k(ctr)$ and a unit vector $u$ perpendicular to $x_A$. Alice then blinds $u$ by using the rotation function $R$, where $\theta$ is pseudorandom value by definition. Alice sends $a = R(u, \theta)$ and $ctr$ to the Server. 

The Server matches the messages having the same $ctr$ value from Alice and Bob, and performs a single inner product operation giving $m = \langle a,b\rangle$, then sends $m$ to Alice. 
\begin{eqnarray} \nonumber
m &=& \langle a,b\rangle  = \langle R(u,\theta),rR(x_B,\theta)+s\rangle  \\ 
&=& r\langle R(u,\theta),R(x_B,\theta)\rangle  + \langle R(u,\theta),s\rangle \label{eqn:m}  
\end{eqnarray}

Since $R$ is an angle preserving map, $R(x_B,\theta)$ is perpendicular to the vector $R(u,\theta)$ if the private values (vectors) of Bob and Alice are same (i.e., $x_A = x_B$). Notice that in this case the inner product on the left of Eqn. (\ref{eqn:m}) vanishes and only $\langle R(u,\theta), s)\rangle$ remains. Alice can compute this value as it does not contain the blinding $r$. Therefore, Alice computes $\langle R(u,\theta), s)\rangle$, and if she finds that $m = \langle R(u,\theta), s)\rangle$, she learns that she has the same private vector as Bob.


\subsection{Approximate and Fake Location Sharing}


At the protocol level, approximate and fake locations are shared in the same manner as actual locations, simply encrypted and routed through the Server using the PSP Protocol. The Albatross system does not guide or make suggestions to the user on what to use as an approximate or fake location. We note, however, that this is an interesting area for future work. Realistic faking especially is ripe with challenges, as the fake locations should be realistic to adversaries with potentially a great deal of auxiliary information. Previous work in faking of locations include~\cite{krumm, wpes, shankar}.     


\subsection{Protocol Chart}
The chart in Table~\ref{tbl:prochart} gives the use of the protocols PSP and VPET with respect to the user's sharing preferences.

\begin{table} [htbp]
\renewcommand\arraystretch{1.2}
\small
\centering
\caption{Albatross' protocol chart.} \label{tbl:prochart}
\begin{tabular}{|@{\ }c@{\ }|@{\ }c@{\ }|@{\ }c@{\ }|@{\ }c@{\ }|@{\ }c@{\ }|@{\ }c@{\ }|@{\ }c@{\ }|} \hline
$A \Rightarrow B$ & available & circle & approx & nearby & invisible & fake \\ \hline
available & PSP & PSP & PSP & PSP & PSP & PSP \\ \hline
circle & PSP & PSP & PSP & PSP & PSP & PSP \\ \hline
approx & PSP & PSP & PSP & PSP & PSP & PSP \\ \hline
nearby & PSP & PSP & PSP & VPET & n/a & PSP \\ \hline
invisible & n/a & n/a & n/a & n/a & n/a & n/a \\ \hline
fake & PSP & PSP & PSP & PSP & PSP & PSP \\ \hline
\end{tabular}
\end{table}

A $\Rightarrow$ B shows how User A shares her location with User B. For instance, the first row is read as available User A shares her location (only) with B using protocol PSP, no matter what User B's status is. 

\section{GPS Data and Grid Systems}
\label{sec:grid}
In this section we give background on GPS coordinates and grid systems for Albatross. We describe a novel grid scheme that provides additional efficiency compared to existing schemes, but at a small reduction in privacy.

\subsection{GPS Coordinates}
When using PSP, one directly shares the location data as captured from the device, say, by the GPS system. In typical applications, acceptable precision can be achieved by packing GPS coordinates (a latitude-longitude pair) into 8 bytes. Hence, in Albatross we allocate 8 bytes for each exact location and truncate input from the GPS system. Table~\ref{tbl:gps} gives the relationship between a given precision and its equivalent physical distance on the earth's surface.

\begin{table}  [htbp]
\renewcommand\arraystretch{1.2}
\centering
\small
\caption{Precision used in GPS data.} \label{tbl:gps}
\begin{tabular}{cc|c|@{\ }c@{\ }|@{\ }c@{\ }|@{\ }c@{\ }|} \cline{3-6}
 & & \multicolumn{4}{|c|}{fractional digits} \\ \cline{3-6}  
\multicolumn{2}{c|}{} & 2 & 3  & 4 & 5\\ \hline
\multicolumn{2}{|c|}{total bits needed} & 31 & 37 & 43 & 51\\ \hline
\multicolumn{2}{|c|}{distance on latitude}  & 1,117 $m$ & 111 $m$ & 11 $m$ & 1.1  $m$ \\ \hline
\multicolumn{1}{|c|}{distance} & polar circles  & 444 $m$ & 44 $m$ & 4.4 $m$  & 0.4 $m$ \\ \cline{2-6}
\multicolumn{1}{|c|}{on} & tropic circles  & 1,021 $m$ & 102 $m$  & 10 $m$  & 1 $m$ \\ \cline{2-6}
\multicolumn{1}{|c|}{longitude} & equator  & 1,113 $m$ & 111 $m$ & 11 $m$  & 1.1 $m$ \\ \hline
\end{tabular}
\end{table}

\subsection{Private Equality Testing and Grids}
\label{sec:petgrids}
In principle, the neighbourhood of a location is defined by a circle and the proximity testing problem is to test whether a contact is inside or outside of the circle. In other words, checking whether a user is nearby or not consists of obtaining the location of the user (e.g., via GPS), calculating the distance to this location, and testing if the distance is lower than some radius $r$.  As pointed out in~\cite{NTL11:Location}, this ideal definition suffers from a triangulation attack that reveals the exact location by stepping across the circular boundary at two different points. Moreover, if the location values are encrypted, the distance calculation would need a costly homomorphic encryption scheme.

\begin{figure}[htbp]
  \centering
    \includegraphics[scale=0.35]{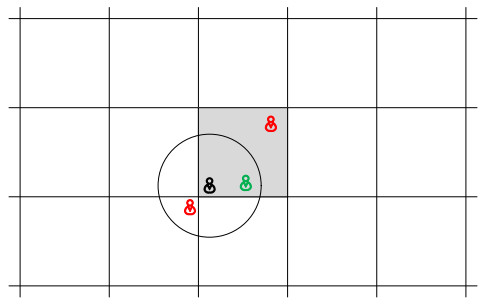}
    \caption{Direct distance computation versus approximation methods. The green user is in both the circle and grid cell and is correctly considered as nearby. However, both of the red users are falsely detected. One of the red users is not in the cell and is thus considered as far away (although he is in the range of the circle). Similarly, the other red user is considered as nearby because he is in the grey cell but he is not in the range of the circle.}
    \label{fig:accuracy}
\end{figure}

Hence, as in~\cite{NTL11:Location}, we reduce private proximity testing to private equality testing (PET), using a grid or tessellation of the surface of the earth. Each grid cell has a unique identifier, and the users can compare the identifiers of the grid cells using a PET protocol. Grid cells do not necessarily match with circular neighbourhoods, so the proximity test is less accurate. For example, in Figure~\ref{fig:accuracy} a user (drawn in black) wants to share his location in the range of a circle while the gray area covers his residing cell.

\begin{figure}[htbp]
  \centering
    \includegraphics[scale=0.45]{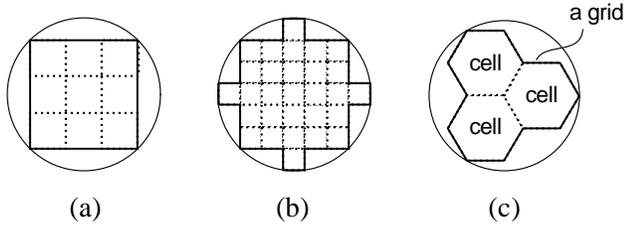}
    \caption{Grids approximating the disk. Observe that (a) needs 9, (b) needs 29, and (c) needs 3 cell comparisons. Moreover (b) is the most and (c) is the least accurate.}
    \label{fig:grid}
\end{figure}

One way to improve the accuracy is to use more and/or smaller cells to approximate the area of the circle as in Figure~\ref{fig:grid}. This approach increases the number of cells to compare, as a user must run a PET protocol to compare grid cell identifiers for each of the cells making up the approximate disk. 

The approximation can be done in different ways, using square cells, hexagonal cells, as well as other shapes, or overlapping layers of cells~\cite{NTL11:Location, NPS12:Location, ZGH07:Louis, MFB11:Privacy}. While all approximation based solutions are inaccurate, certain shapes can approximate a circle using fewer cells than others.

\subsection{Albatross Grids and Cells}
The Albatross system does not depend on any particular grid scheme. Nevertheless, we have chosen to construct our own grid scheme which is highly efficient (with only one PET protocol run) but with a potential reduction in privacy compared to other schemes. To describe our scheme, we deviate from standard terminology slightly. In the literature, a grid cell means an element of the grid tessellation. In this paper, we say the grid is the tessellation of the surface of the earth into approximate disks. Each element of the grid is further subdivided into cells. A grid element is the union of cells approximating the circle. 


\begin{figure}[htbp]
  \centering
    \includegraphics[scale=0.45]{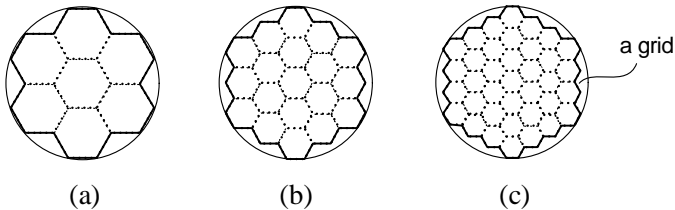}
    \caption{Grids with hexagonal cells.}
    \label{fig:hexgrid}
\end{figure}

In order to approximate the area of the circle with cells, each location can be tagged with a grid ID and a cell ID. Assume that the grid consists of a collection of cells that cover the location space without overlap. In other words, grids have a regular shape such that there is no location left uncovered. For instance, grids such as (a) and (c) in Figure~\ref{fig:grid} can cover the location space without overlap but not (b) because of its non-regular boundary.


Accuracy of a circle covering approximation can be increased by increasing the number of cells in a grid. This can be done effectively using hexagonal cells as seen in Figure~\ref{fig:hexgrid}. Observe that (a) has 7, (b) has 19, and (c) has 37 cells in their grids. As the number of cells increases, accuracy increases. Note also that all these grids span the location space evenly when tiled correctly.  

\begin{figure*}[t]
\centering
\subfloat[][]
{\includegraphics[scale=0.7]{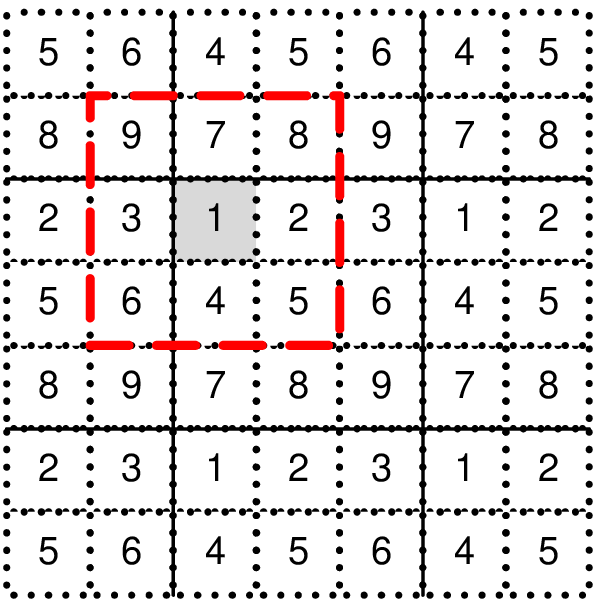}
\label{fig:latinsquare}}
\qquad\qquad
\subfloat[][]
{
\includegraphics[scale=0.44]{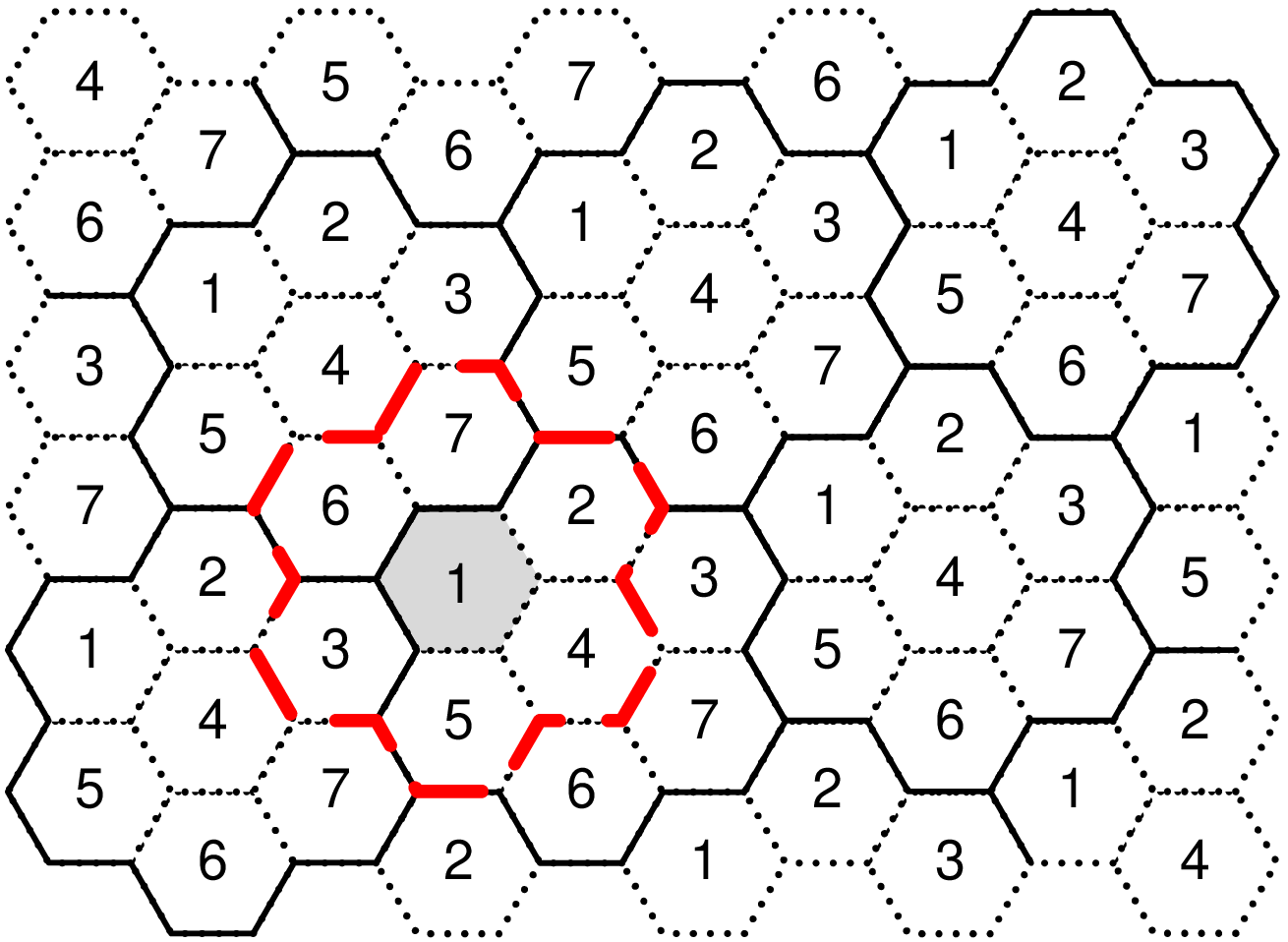}
\label{fig:hexlabel}
}
\caption{Cells are labeled so that any approximate disk is a ``Latin Square''. For a grid of square elements composed of 9 cells each, any approximate disk has 9 cells with exactly one cell labeled from 1 to 9. For grid elements composed of 7 hexagons, any approximate disk having 7 cells has exactly one cell labeled from 1 to 7.}
\label{fig:celllabels}
\end{figure*}

\subsection{Canonical Grid Identifiers}
We describe here how the whole globe might be mapped by a grid system. The grid granularity must first be fixed. This corresponds to a notion of how to quantify when two users are considered ``nearby''. 

\begin{figure}[htbp]
  \centering
    \includegraphics[scale=0.6]{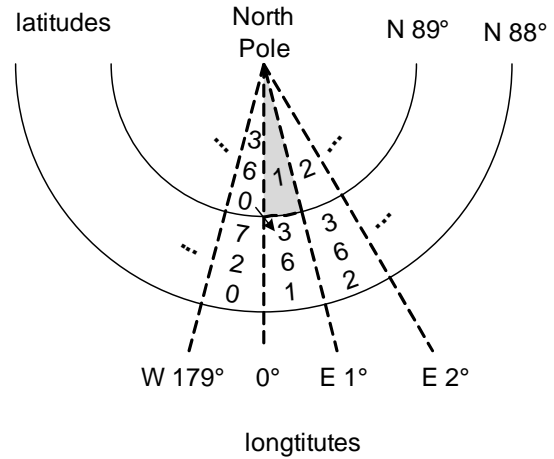}
    \caption{A simple labeling of grid elements.}
    \label{fig:gridID}
\end{figure}

{\bf Grid Element labels:} Starting from the north pole and the 0th longitude passing from Greenwich, one can enumerate the grids having, say, 1 degree width and length until reaching the south pole as seen in Figure~\ref{fig:gridID}. Note that in this grid system, the grid elements are not perfect squares but for simplicity we treat them as squares. The grid elements then can be uniquely labeled with a simple enumeration. 

{\bf Cell labels:} Each cell in a grid element is labeled with an identifier that is unique within the grid element. Moreover, this labeling is done so that across the whole location space each cell's identifier is unique in an approximate disk centered at that cell. For instance, in Figure~\ref{fig:celllabels} we give two examples of cell labels for a portion of the whole plane.

\subsection{Grid Proximity Protocol}
We present the protocol used in proximity testing for Albatross. Our method needs only a single comparison (i.e., VPET protocol run), independent of the number of approximation cells. The protocol starts as with a grid and cell system labeled as in the previous section. One party provides a cell label to the other party, and both parties are able to select grid elements for equality testing. The two parties are nearby if and only if the grid elements are equal.

\begin{figure}[htbp]
  \centering
    \includegraphics[scale=0.5]{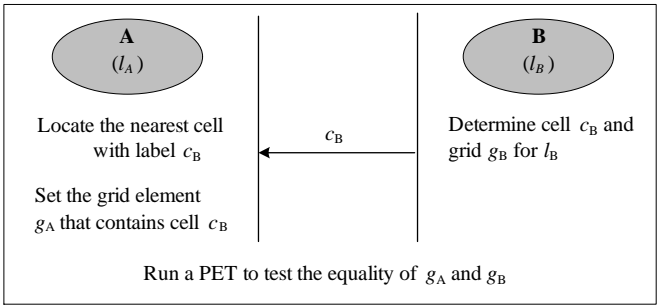}
    \caption{Grid Proximity Protocol.}
    \label{fig:compare}
\end{figure}

Figure~\ref{fig:compare} shows the steps of the protocol. Let $l_A$ and $l_B$ represent the exact locations of User A and User B, respectively. First, User B computes the cell $c_B$ and grid element $g_B$ corresponding to his location. User A locates the nearest cell with label $c_B$. User A selects the grid element containing this cell, call is $g_A$. User B uses the grid element $g_B$ in the VPET protocol run with User A. User A compares $g_B$ to see if it is equal to $g_A$.

\begin{figure}[htbp]
  \centering
    \includegraphics[scale=0.5]{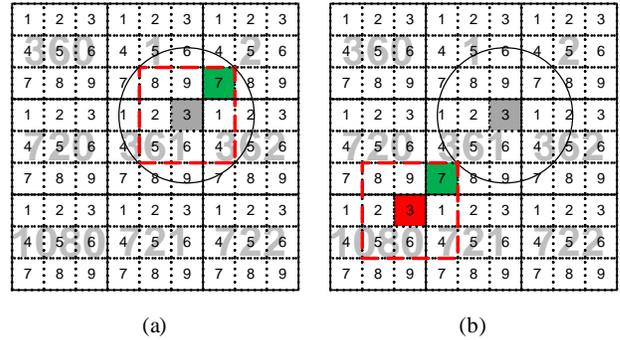}
    \caption{Grid Protocol Example.}
    \label{fig:GridProtocolExample}
\end{figure}

In order to describe the process more clearly, we work through the protocol in a toy example. We first consider the case where two users are nearby (Figure~\ref{fig:GridProtocolExample}a). Suppose User A wants to learn whether B is nearby or not. User B computes his grid element to be "2" and cell label as "7". Nearby thus corresponds to be inside the circle in Figure~\ref{fig:GridProtocolExample}a, approximated by the red-colored square. User B sends the cell label 7 to User A. After User A receives User B's cell label, User A takes the cell nearest to his location with the same label. By virtue of the Latin Square property, there is only one such cell. User A knows that if User A and User B are nearby, User B should be in this cell. Therefore, User A looks up the grid element containing this cell and determines that it is "2". B then sends this grid element to A in the VPET protocol. After a run of VPET, User A concludes that User B is nearby, although they reside in different grids in the grid system.  

Let us change the location of User B slightly and repeat the same experiment. Assume that User A is in the same location but User B is located at cell 7 in grid 361 (Figure~\ref{fig:GridProtocolExample}b).
After User A gets User B's cell label, User A assumes that User B should be in the closest cell labeled with 7 (i.e., the red cell in Figure~\ref{fig:GridProtocolExample}b) and sends the respective grid element "2" to A in the VPET protocol comparison. Since User B's grid element is "361", she concludes that they are not nearby.

%

Note that the grid system in use can be fixed or can be negotiated. To the extreme, any GPS location can be viewed as a cell and according to the proximity needed a hexagon approximating a circle can be used for the equality testing.

Note that our system trades efficiency for a reduction in privacy for User B. User A gains knowledge of User B's cell label. If User B has little idea of User A's whereabouts, the leakage of the cell label does little harm since there are many cells with the same label. However, if User B has auxiliary knowledge about User A's location, then knowing the cell label can further narrow down User A's location. In this way, our system shares a weakness of existing work: the proximity protocols leak more than strictly desired. If User B is actually ``nearby'', User A can localize User B to a smaller area than an approximate disk around User A. In (a) of Figure~\ref{sec:grid}, the 9 comparisons would reveal which individual cell User B is in. This is an interesting area for future work (see the Appendix in~\cite{NTL11:Location} for an approach to this problem).

\section{Unified Location Protocol}
\label{sec:unified}
As described in Section~\ref{sec:AlbatrossPrivacyGoals}, the active social network of a user is considered private information. From the sharing habits of a user, one can determine the user's circles and close relationships. One of goals of Albatross is to hide this information from the server and other users. To accomplish this goal, we use protocol hiding by unifying base protocols.

To explain the problem in more detail, consider Table~\ref{tab:ServerViewNotUnified} where we give an example of the sort of information that can be inferred by the server. One problem is that the PSP and VPET protocols differ in structure. Hence, by virtue of routing the traffic, the server can determine whether User A's location sharing granularity for User B is ''nearby'' or not. This violates Property P1 of our privacy goals.
\begin{table} [htbp]
\renewcommand\arraystretch{1.2}
\small
\centering
\caption{Server view without unified protocol. The server can see whether a user's preference for other users is nearby or not. The server can also see contacts a user is not sharing with and which users are offline.}
\begin{tabular}{|@{\ }c@{\ }|@{\ }c@{\ }|@{\ }c@{\ }|@{\ }c@{\ }|@{\ }c@{\ }|@{\ }c@{\ }|@{\ }c@{\ }|} \hline
         & User $A$ & User $B$ & User $C$ & User $D$ & User $E$ & $\hdots$ \\ \hline 
User $A$ & n/a      & PSP      & invis.   & VPET      &    PSP & $\hdots$      \\ \hline 
User $B$ & PSP      & n/a      & invis.   & VPET      &  PSP & $\hdots$      \\ \hline 
User $C$ & VPET     &  PSP   & n/a     & VPET      & PSP & $\hdots$    \\ \hline 
User $D$ & PSP      &  PSP     & PSP    & n/a     & PSP & $\hdots$   \\ \hline
User $E$ & offl.    &  offl.   & offl.    & offl.    & n/a & $\hdots$  \\ \hline
$\vdots$ & $\vdots$ &  $\vdots$ & $\vdots$  & $\vdots$ & $\vdots$ & $\ddots$  \\ \hline 
\end{tabular}
\label{tab:ServerViewNotUnified}
\end{table}

To address this problem, we describe a Unified Location Protocol which masks to the server which protocol (PSP or VPET) is being used. This protocol is asynchronous, like its underlying base protocols. This protocol is used with all contacts. If the preference is to share nothing (i.e., remain invisible), an agreed upon dummy location can be shared.

However, another problem then appears: User A may choose to be invisible to User B, which means User A does not share location with User B. This preference is supposed to be hidden from the server, so A must still checkin to B, or else the server would know the preference. However, User B can then tell whether User A is offline or just being invisible with respect to User B (see Table~\ref{tab:UserViewNoCaching}) as presumably User A must be online in order to checkin. This problem is more apparent with periodic location checkins or ``tracking'' as opposed to sporadic checkins. This violates Property P2 of our privacy goals.
\begin{table} [htbp]
\renewcommand\arraystretch{1.2}
\small
\centering
\caption{User A's view without caching. Besides the sharing preferences of his contacts, User A can also see whether each contact is offline, if they checkin periodically.}
\begin{tabular}{|c|c|c|c|c|c|} \hline
  &  User $B$ &  User $C$ &  User $D$ &  User $E$ & $\dots$\\ \hline 
status   &   nearby & invisible & available & offline & $\dots$ \\ \hline 
\end{tabular}
\label{tab:UserViewNoCaching}
\end{table}

Property P2 says that if a user goes offline, his contacts should not be aware of this fact. We address this problem with a caching scheme used in conjunction with the Unified Location Protocol. The user employs the caching scheme with the help of the server. The server will be aware that a user is offline, and can help the user maintain the illusion of still being active to his contacts.


\subsection{Protocol Masking}
By masking the protocol traffic, we hide the user's active social network from the server. For instance, the server cannot tell that User A is hiding from User B, as opposed to some other location sharing preference, such as ``Available''.

\begin{figure}[htbp]
\centering
\subfloat[][\small PSP protocol]{
\centering
    \includegraphics[scale=0.3]{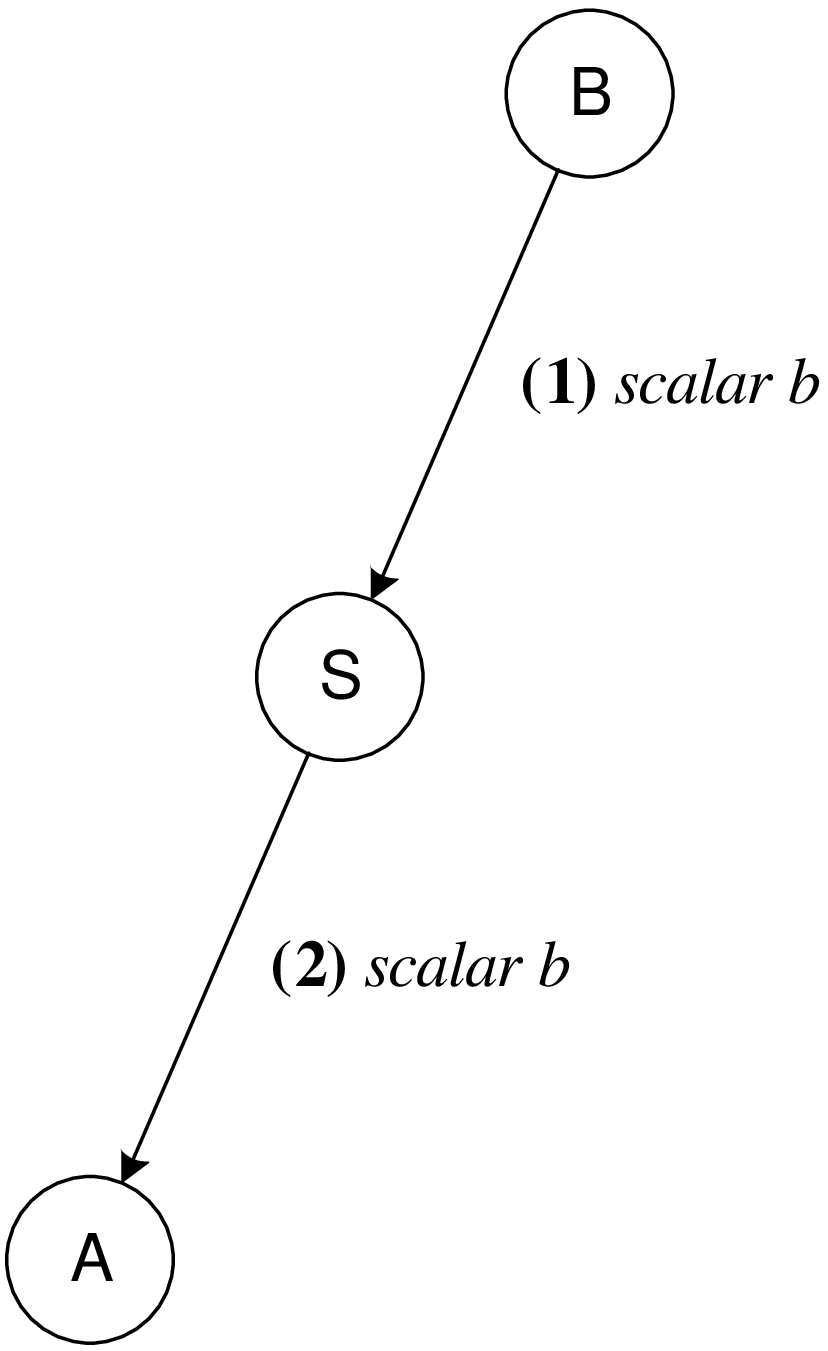}
    \label{fig:PSP_flow}
}
\qquad \qquad 
\subfloat[][\small VPET protocol]
{
  \centering
    \includegraphics[scale=0.3]{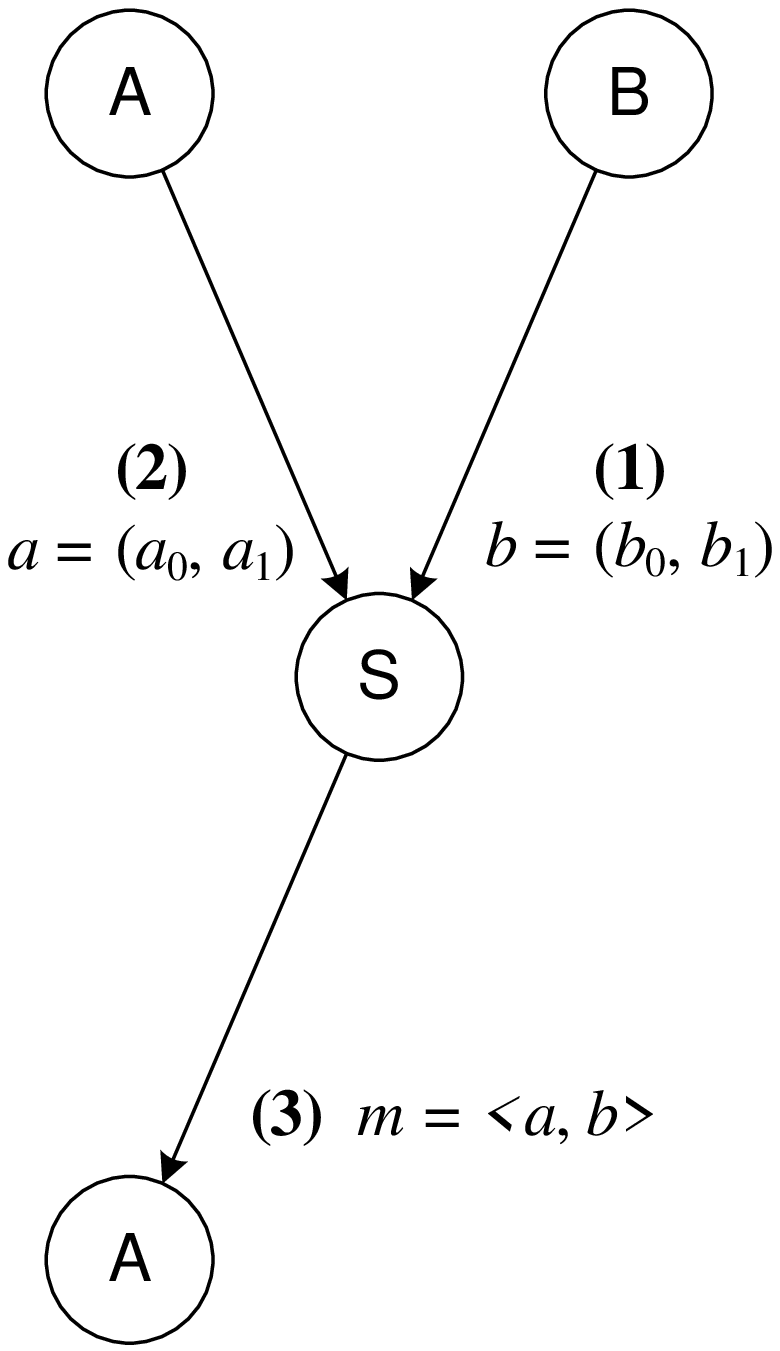}
    \label{fig:VPET_flow}
}

\subfloat[][\small Unified protocol]
{
\centering
    \includegraphics[scale=0.3]{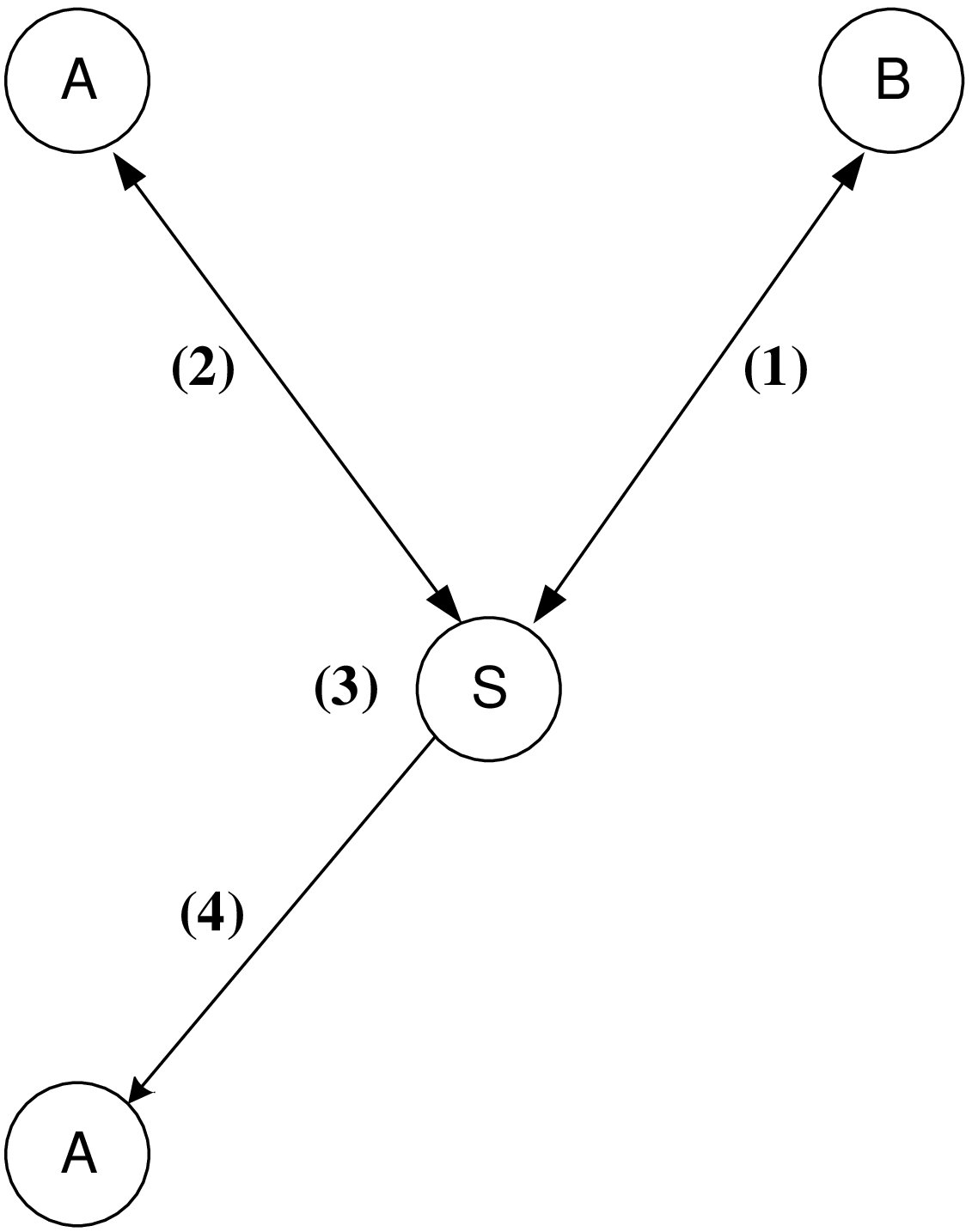}
    \label{fig:unified_flow}
}
\caption{Unified Protocol encapsulating both PSP and VPET protocols}
\label{fig:proto_flows}
\end{figure}

We overload specific location coordinates to carry special meaning: $x_n$ (= not nearby), $x_y$(= nearby), and $x_i$ (= invisible). These are {\it dummy} location coordinates and used to hide communication from the server through the PSP protocol. These coordinates, for instance, have longitudes larger than 180. The $x_i$ coordinate is used with PSP when a user wants to be invisible with respect to another user.

Figure~\ref{fig:proto_flows} shows the structure of the PSP and VPET protocols; the structure of the Unified Protocol simply encapsulates both base protocols. There are two protocol actions associated with the Unified Protocol: Checkin and Retrieval. Checkin is performed by User B to share location with User A. Retrieval is performed by User A to see User B's location. Retrieval can be performed anytime after checkin, even after User B is offline, due to the asynchronous nature of the protocol.

As described in Section~\ref{sec:protocols}, we assume A and B share a key $k_{AB} $ and they can create keys $k_1$ and $k_2$ by encrypting the $ctr$ value with $k_{AB} $. In the same way, a user can create a random bit to encrypt one bit of information.

\vspace{0.3cm}
\noindent{\bf \underline{Checkin: User B shares with User A}}

\noindent Before checkin, User B may have done a retrieval of A's location information at some point in the past. This information is used in the case that A's preference is PSP and B's preference is VPET. If B has no information on A's location, B assumes A is ``invisible.'' 

\vspace{0.3cm}
\noindent{\bf Step 1:} User B generates a random number for the $ctr$ value. He encrypts $ctr$ with $k_{AB} $. He places the $ctr$ value on the server and two encrypted bits. One bit is the protocol bit for his sharing granularity towards A, representing whether B wants to use the VPET or PSP protocol with A, i.e., whether B's location sharing preference for User A is ``nearby'' or not. This is $b_{B\rightarrow A}$. The other bit is A's sharing granularity towards B, stored from the last retrieval of A: $b_{A\rightarrow B}$. If his sharing preference is ``nearby'' he also places his cell number.

\begin{table} [htbp]
\renewcommand\arraystretch{1.2}
\small
\centering
\caption{Step 1 of Unified Protocol. User B stores these values on server.}
\begin{tabular}{|c|c|} \hline
Field                              & Description \\ \hline\hline
counter                            & mandatory  \\ \hline 
Protocol bit for $B\rightarrow A$: $b_{B\rightarrow A}$  & mandatory \\ \hline
Protocol bit for $A\rightarrow B$: $b_{A\rightarrow B}$  & random bit if $b_{B\rightarrow A}$ is PSP \\ \hline 
cell number                        & random if $b_{B\rightarrow A}$ is PSP \\ \hline
B's vector                         & mandatory \\ \hline
\end{tabular}
\label{tab:ProtStep1}
\end{table}

\noindent There are three cases:\\
{\bf A is VPET/B is VPET}: B proceeds as in the VPET protocol and sends to the server $v_2 = r R(x_B, \theta) + s$.\\
{\bf A is PSP/B is VPET}: From B's last retrieval of A's location, B computes and sends to the server the encrypted location vector $(x_y \bigoplus k_1, k_2)$ or $v_2 = (x_n \bigoplus k_1, k_2)$ depending on whether A is nearby or not.\\
{\bf Otherwise}: B sends to the server the vector $(x_B \bigoplus k_1, k_2)$. B may send $x_i$ for $x_B$ if he wants to be invisible to A.

\vspace{0.3cm}
\noindent{\bf \underline{Retrieval: User A gets User B's location information}}

\noindent{\bf Step 2:} A retrieves $ctr$ and $b_{B\rightarrow A}$ and $b_{A\rightarrow B}$. If $b_{B\rightarrow A}$ is VPET and $b_{A\rightarrow B}$ is out-of-date, A aborts. Otherwise, there are two cases:\\
{\bf A is VPET/B is VPET}: A proceeds as in the VPET protocol and sends the vector $v_1 = R(u,\theta)$ to the server.\\
{\bf Otherwise}: A sends a random vector $(b_1, b_2)$, one that corresponds to a location through the vectorization process. We cannot use a completely random vector as this would be distinguishable from the vector in the other case.

\vspace{0.3cm}
\noindent\hangindent=0.3cm {\bf Step 3:} The server computes the vector inner product $m = \langle v_1, v_2 \rangle$ and sends the result to A.

\vspace{0.3cm}
\noindent\hangindent=0.3cm {\bf Step 4:} A decrypts inner product. There are two cases:\\
{\bf A is VPET/B is VPET}: As in the VPET protocol, A computes $\langle R(u,\theta), s)\rangle$ and if this quantity is equal to $m$, then B is nearby.\\
{\bf Otherwise}: A receives $m = b_1(x_B \bigoplus k_1) + b_2 k_2$. A knows all values except $x_B$ and so can compute $x_B$.

\vspace{0.3cm}
\noindent Notes:\\
(1) In all cases Server does the same operation in Step 3, an inner product, and cannot infer anything from $v_1$ and $v_2$.\\
(2) The unified protocol as described only shows B sharing with A. In reality, there is also A sharing with B. Both directions have their own counters, keys, and preferences.

\subsection{Counter Caching}
Since the server is assumed to be honest-but-curious, there is no need to defend against it performing message-replay attacks.  By allowing the server to replay messages, we enable it to reissue advertisements of invisibility on behalf of clients who are actually offline. In other words we allow the server to help the user maintain the illusion of being online to other users. The idea is that Step 1 can be done multiple times by User B in batch, each time generating a new counter and specifying the PSP protocol for B's granularity towards A. User B can also do Step 3 in batch, specifying $x_i$ (=invisible) for $x_B$ and sending $v = (x_B \bigoplus k_1, k_2)$. The server keeps a conceptual table for each user, as in Table~\ref{tab:caching}. At the end of the protocol, User A sees that User B is invisible and cannot tell if User B is actually online or not. After a counter value has been used, the corresponding row in the table can be deleted.

\begin{table} [htbp]
\renewcommand\arraystretch{1.2}
\small
\centering
\caption{User's counter cache. This information is used by the server to help the user checkin even when offline and thus maintain the illusion of being online. The checkins the server makes on behalf of the user corresponds to the ``invisible'' preference; hence all contacts will see this user as ``invisible.''}
\begin{tabular}{|c|c|c|c|c|c|} \hline
Counter & Protocol bit & Encrypted location \\ \hline 
ctr1    & E(PSP) & $v_1$ \\ \hline 
ctr2    & E(PSP) & $v_2$  \\ \hline 
ctr3    & E(PSP) & $v_3$  \\ \hline 
$\vdots$ & $\vdots$ & $\vdots$ \\ \hline 
\end{tabular}
\label{tab:caching}
\end{table}

If the user is actually online and wishes to change his sharing granularity, he can generate a new counter value and prepend a new row to his counter cache table.

\subsection{Privacy Analysis with Unified Protocol}
In Tables~\ref{tab:ServerViewUnified} and~\ref{tab:UserViewUnified} we see the end result of our unified protocol. The server sees only whether the user is on or off. The user sees only the sharing granularity of his contacts towards himself, not whether they are online or offline.

\begin{table} [htbp]
\renewcommand\arraystretch{1.2}
\small
\centering
\caption{Server view with unified protocol. The server sees only whether a user is online or not, as well as the set of contacts for each user. Nothing about the preferences for each contact can be seen by the server.}
\begin{tabular}{|c|c|c|c|c|c|} \hline
  &  User $B$ &  User $C$ &  User $D$ &  User $E$ & $\dots$\\ \hline 
status   &   on & on & on & off & $\dots$ \\ \hline 
\end{tabular}
\label{tab:ServerViewUnified}
\end{table}

\begin{table} [htbp]
\renewcommand\arraystretch{1.2}
\small
\centering
\caption{User A's view with unified protocol and caching. User A can see the preferences of his contacts towards User A, but cannot tell if users are invisible or simply offline. In this way, his contacts maintain plausible deniability about being offline versus purposefully hiding from User A.}
\begin{tabular}{|c|c|c|c|c|c|} \hline
  &  User $B$ &  User $C$ &  User $D$ &  User $E$ & $\dots$\\ \hline 
status   &   nearby & invisible & available & invisible & $\dots$ \\ \hline 
\end{tabular}
\label{tab:UserViewUnified}
\end{table}

\section{System Prototype} \label{sec:sys}
Albatross is realized as a working research prototype and we describe the implementation here.
With this prototype, we demontrate that the performance of Albatross is suitable for real-world deployment and raises no scalability concerns. The overhead compared to a non-private system is on the order of the average number of contacts.

\begin{figure}[htbp]
  \centering
    \includegraphics[scale=0.65]{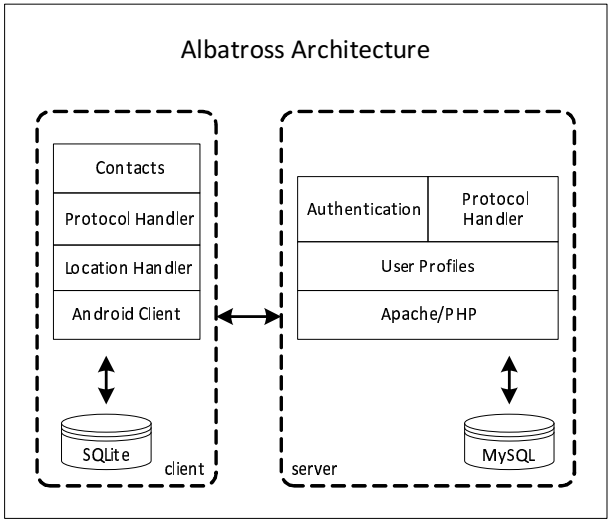}
    \caption{High-level Albatross architecture. The server is built on top of Apache/PHP/MySQL. The client is an Android application with persistent storage in SQLite.}
    \label{fig:unified}
\end{figure}

\subsection{Server-side}
We used Apache/PHP and MySQL as the backbone of our server-side implementation (see Fig. \ref{fig:unified}). The main MySQL tables are a users table that contains profile data and a contacts table that contains a list of contacts (i.e., edges in the social graph) in the system. Each edge in the contacts table contains protocol information for both protocol directions, for example, User A's sharing with User B and also User B's sharing with User A. Registration and authentication is through a username and password.

\subsection{Client-side}
We developed an Android client for Albatross. This client application keeps sharing preferences and keys for each contact in the native Android SQLite database. SQLite gives us a self-contained SQL database to store contact data even after Albatross is stopped.

The client consists of a protocol handler that determines the protocol to use and performs the Unified Protocol steps. The protocol handler uses the Spongy Castle cryptographic library, a repackaging of the standard Bouncy Castle cryptographic libraries explicitly for the Android platform. Albatross employs Spongy Castle for AES and Diffie-Hellman key exchange calculations. 

Note that we relied on a public/private key pair on the Android devices in the Diffie-Hellman key exchange. These key pairs were generated as part of Albatross installation. The client also contains a location handler which gets location information from the device via the GPS system. Corresponding to the GPS location, the protocol handler determines the grid and cell values if using the VPET protocol.


\subsection{Performance}
We measured the total time for a retrieval on a Samsung Galaxy S III smartphone having a 1.5 GHz dual-core Qualcomm Krait processor running Android 4.0.4. The time for a checkin should be even less. Our server was an Ubuntu 12.04 LTS instance in the Amazon Elastic Cloud (EC2), running on an AMD 64 with 1.7 GB of memory. As our server was dedicated to only Albatross, we picked a small instance (m1.small) providing a single EC2 compute unit.  

\begin{figure}[htbp]
  \centering
    \includegraphics[scale=0.5]{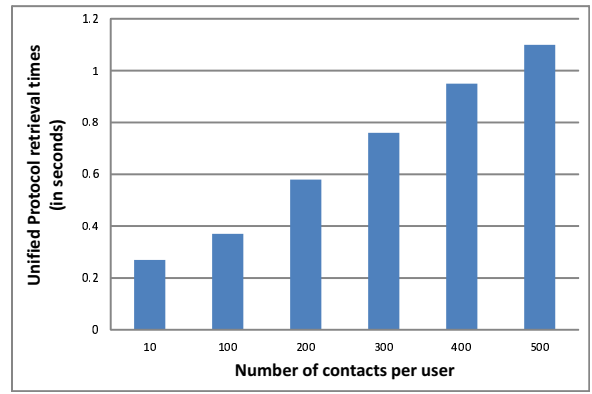}
    \caption{Retrieval time for various numbers of contacts.}
    \label{fig:timing}
\end{figure}

Recall that in the Unified Protocol, users share some location -- real, approximate, fake or dummy -- with all of their contacts. The Unified Protocol has two halves, a checkin and a retrieval. For experimental purposes, we placed checkin data on the server and executed the Unified Protocol for retrieval (the more complex and time-sensitive half of the protocol) in batch. Location information for all contacts was downloaded, corresponding to a user finding where all his friends are. We assumed that users have around 500 contacts total. Averaging over 20 retrievals, we found that it took about 1.1 seconds for a user to retrieve location information for all 500 contacts. In order to see the effect of the number of users, we repeated our experiments for various numbers of contacts as illustrated in Figure~\ref{fig:timing} and the time scales about linearly with the number of contacts.   

One question is how our privacy requirements effect the total amount of storage and user/server workloads. Compared to a simple location uploading for a trusted server, as in Google Latitude and Foursquare, Albatross requires protocol interaction with each contact (through the server) rather than an interaction with just the server. Also, the Unified Protocol is more complex than a simple upload. In Tables~\ref{tab:perf} and~\ref{tab:storage} we give a theoretical description of the performance and storage costs of the Unified Protocol in terms of $N$, the number of a user's contacts. For a trusted server, each checkin is simply the uploading of location information to the server. For Albatross, each checkin involves one symmetric encryption and random number generation on the client side for each of the user's contacts. Each retrieval requires two client-server interactions. Since the number of contacts in a social network generally runs in the several hundreds~\cite{PEW_Social}, as a rough estimate the performance and storage costs for Albatross can be on the order of $N$ times more than for a non-private system.

\begin{table} [htbp]
\renewcommand\arraystretch{1.2}
\small
\centering
\caption{Summary of Albatross computation costs for one user's checkin and retrieval. In Albatross, checkins must be performed per contact. The checkin is also more involved, with a symmetric encryption and random number generation per contact. The retrieval using Albatross involves two client-server interactions, the server computes a different inner product for each contact, and each inner product must be decrypted. Hence, the computation costs for Albatross scale with the number of contacts $N$. Below, $C$ represents a constant cost independent of $N$.}
\label{tab:perf}
\begin{tabular}{|c||c|c|c|} \hline
 & Client cost & Server & Client cost \\
Protocol & (Retrieval) & cost & (Checkin) \\ \hline \hline
Unified     & $O(N)$  & $O(N)$ & $O(N)$ \\ \hline
Non-private  & C  & C   & C \\ \hline
\end{tabular}
\end{table}

\begin{table} [htbp]
\renewcommand\arraystretch{1.2}
\small
\centering
\caption{Albatross storage requirements per user. In Albatross, clients store a key (16 bytes) and location preference (1 byte) for each contact. Servers store state information for the Unified Protocol (17 bytes, see Table~\ref{tab:ProtStep1}) and also, say, 10 cache values of state information. In a non-private, trusted server protocol, servers store the user's location and location preferences per contact.}
\label{tab:storage}
\begin{tabular}{|c||c|c|} \hline
         & Client  & Server \\
Protocol & storage & storage \\ \hline \hline
Unified Protocol     & $(16+1)N$  & $17*11*N$ bytes \\ \hline
Non-private Protocol & 0    & $8 + N$ bytes \\ \hline
\end{tabular}
\end{table}

\section{Related Work}
Existing production location sharing systems with a trusted server include Foursquare~\cite{Foursquare}, Google Latitude~\cite{Latitude}, Apple's Find My Friends~\cite{FindMyFriends}, and Glympse~\cite{glympse}. There has been work on location sharing systems that enhance privacy through an untrusted server. The work by Narayanan et al.~\cite{NTL11:Location} concentrates on a particular type of location sharing, the ``nearby'' sharing granularity. We aim for a more complete system that includes different granularities of location sharing and address the issue of how to hide these granularities from the server. Zhong et al.~\cite{ZGH07:Louis}, Mascetti et al.~\cite{MBF09:Longitude, MFB11:Privacy}, and Nielsen et al.~\cite{NPS12:Location} also focus on the ``nearby'' sharing granularity.  
 
Private versions of other types of social networking systems with an untrusted server have also been studied. Cristofaro et al.~\cite{Hummingbird} studied micro-blogs, specifically Twitter. Their system is analogous to ours in that they attempt to provide the expected core micro-blogging services, and yet hide tweets, hashtags, and followers from the server. Rieffel et al.~\cite{rieffel} study presence systems, in which a user's presence status (``in building'', ``in office'', ``has visitor'', etc.) is hidden from the server. There is a large body of work studying recommender systems with untrusted servers; early work in this area was performed by Canny~\cite{canny}.

Location privacy in general is a well-studied topic. Problems such as anonymity and obfuscation of location data is one area of concentration. Krumm~\cite{KrummSurvey} provides a good survey for this area. Another area of study is user preferences for location sharing, for instance, see~\cite{Toch, Benisch}.

\section{Conclusion} \label{sec:conc}
We present Albatross, a new privacy-enhanced location sharing service. We designed a complete system with a full selection of location sharing granularities. Such a complete system may reveal social networks, i.e., individual social circles, through analysis of individual sharing granularities over time. Previous work has in general concentrated on a particular sharing granularity, such as ``nearby''.

In our design of Albatross we analyse desired privacy properties of a location sharing service and achieved location privacy from the service provider and social network privacy. The level of privacy attained is not complete, but seems reasonable given the practical considerations of keeping traffic at a reasonable level. We achieved location privacy through lightweight cryptographic protocols and social network privacy through protocol unification and masking. A working research prototype of Albatross showed performance is acceptable.

There are numerous opportunities to improve and extend Albatross. For example, privacy may be further improved by protecting a user's offline/online patterns from the server through a proxy architecture. Also, we plan to investigate how a system such as Albatross can support services such as advertising and location recommendation. After all, the appeal of holding consumer location information is the potential increase in precision and relevance of recommendations and targeted ads. We believe advertising and location recommendation systems are still possible with Albatross but of course will have to be re-designed.


\begin{thebibliography}{10}

\bibitem{albatross}
Albatross -- wikipedia, the free encyclopedia.

\bibitem{FacebookLocation}
Facebook is said to create mobile location-tracking app.
\newblock
  http://www.bloomberg.com/news/2013-02- 04/facebook-is-said-to-create-mobile-location-tracking- app.html.

\bibitem{FindMyFriends}
Find my friends.
\newblock http://www.apple.com/icloud/features/find-my- friends.html.

\bibitem{Foursquare}
Foursquare.
\newblock \url{https://foursquare.com/about/}.

\bibitem{glympse}
Glympse.
\newblock \url{http://www.glympse.com/}.

\bibitem{Latitude}
Google latitude.
\newblock \url{http://www.google.com/mobile/latitude/}.

\bibitem{Highlight}
Highlight.
\newblock \url{http://highlig.ht/}.

\bibitem{AD01:Secure}
M.~J. Atallah and W.~Du.
\newblock Secure multi-party computational geometry.
\newblock In {\em Algorithms and Data Structures, 7th International Workshop,
  WADS 2001}, volume 2125 of {\em Lecture Notes in Computer Science},
  Providence, RI, USA, 2001. Springer.

\bibitem{Benisch}
M.~Benisch, P.~G. Kelley, N.~Sadeh, and L.~F. Cranor.
\newblock Capturing location-privacy preferences: quantifying accuracy and
  user-burden tradeoffs.
\newblock {\em Personal Ubiquitous Comput.}, 15(7):679--694, Oct. 2011.

\bibitem{PEW_Privacy}
J.~L. Boyles, A.~Smith, and M.~Madden.
\newblock Privacy and data management on mobile devices.
\newblock
  \url{http://pewinternet.org/},
  2012.

\bibitem{canny}
J.~Canny.
\newblock Collaborative filtering with privacy.
\newblock In {\em Proceedings of the IEEE Symposium on Security and Privacy},
  2002.

\bibitem{wpes}
R.~Chow and P.~Golle.
\newblock Faking contextual data for fun, profit, and privacy.
\newblock In {\em WPES}, pages 105--108, 2009.

\bibitem{Hummingbird}
E.~D. Cristofaro, C.~Soriente, G.~Tsudik, and A.~Williams.
\newblock Hummingbird: Privacy at the time of twitter.
\newblock In {\em IEEE Symposium on Security and Privacy}, pages 285--299,
  2012.

\bibitem{FNW96:Comparing}
R.~Fagin, M.~Naor, and P.~Winkler.
\newblock Comparing information without leaking it.
\newblock {\em Communications of the ACM}, 39:77--85, 1996.

\bibitem{PEW_Social}
K.~N. Hampton, L.~S. Goulet, L.~Rainie, and K.~Purcell.
\newblock Social networking sites and our lives.
\newblock
  \url{http://www.pewinternet.org/},
  2011.

\bibitem{JY96:Proving}
M.~Jakobsson and M.~Yung.
\newblock Proving without knowing: On oblivious, agnostic and blindfolded
  provers.
\newblock In {\em Proceedings of the 16th Annual International Cryptology
  Conference on Advances in Cryptology}, CRYPTO '96, pages 186--200, London,
  UK, UK, 1996. Springer-Verlag.

\bibitem{krumm}
J.~Krumm.
\newblock Realistic driving trips for location privacy.
\newblock In {\em Pervasive}, pages 25--41, 2009.

\bibitem{KrummSurvey}
J.~Krumm.
\newblock A survey of computational location privacy.
\newblock {\em Personal Ubiquitous Comput.}, 13(6):391--399, Aug. 2009.

\bibitem{MBF09:Longitude}
S.~Mascetti, C.~Bettini, and D.~Freni.
\newblock Longitude: Centralized privacy-preserving computation of users'
  proximity.
\newblock In {\em Proceedings of the 6th VLDB Workshop on Secure Data
  Management}, SDM'09, pages 142--157, Berlin, Heidelberg, 2009.
  Springer-Verlag.

\bibitem{MFB11:Privacy}
S.~Mascetti, D.~Freni, C.~Bettini, X.~S. Wang, and S.~Jajodia.
\newblock Privacy in geo-social networks: proximity notification with untrusted
  service providers and curious buddies.
\newblock {\em The VLDB Journal}, 20(4):541--566, Aug. 2011.

\bibitem{NTL11:Location}
A.~Narayanan, N.~Thiagarajan, M.~Lakhani, M.~Hamburg, and D.~Boneh.
\newblock Location privacy via private proximity testing.
\newblock In {\em Proceedings of the Network and Distributed System Security
  Symposium (NDSS), San Diego, California, USA}. The Internet Society, 2011.

\bibitem{NPS12:Location}
J.~D. Nielsen, J.~I. Pagter, and M.~B. Stausholm.
\newblock Location privacy via actively secure private proximity testing.
\newblock {\em Pervasive Computing and Communications Workshops, IEEE
  International Conference on}, 0:381--386, 2012.

\bibitem{rieffel}
E.~G. Rieffel, J.~T. Biehl, B.~van Melle, and A.~J. Lee.
\newblock Secured histories for presence systems.
\newblock In {\em CTS}, pages 446--456, 2011.

\bibitem{SCJ13:Private}
G.~Saldamli, R.~Chow, H.~Jin, and B.~Knijnenburg.
\newblock Private proximity testing with an untrusted server.
\newblock In {\em Proceedings of the Sixth ACM Conference on Security and
  Privacy in Wireless and Mobile Networks}, WiSec '13, pages 113--118, New
  York, NY, USA, 2013. ACM.

\bibitem{shankar}
P.~Shankar, V.~Ganapathy, and L.~Iftode.
\newblock Privately querying location-based services with sybilquery.
\newblock In {\em Proceedings of the 11th international conference on
  Ubiquitous computing}, Ubicomp '09, pages 31--40, New York, NY, USA, 2009.
  ACM.

\bibitem{S12:46}
A.~Smith.
\newblock 46\% of american adults are smartphone owners.
\newblock Technical report, Pew Research Center, March 2012.

\bibitem{Toch}
E.~Toch, J.~Cranshaw, P.~H. Drielsma, J.~Y. Tsai, P.~G. Kelley, J.~Springfield,
  L.~F. Cranor, J.~I. Hong, and N.~M. Sadeh.
\newblock Empirical models of privacy in location sharing.
\newblock In {\em UbiComp}, pages 129--138, 2010.

\bibitem{ZGH07:Louis}
G.~Zhong, I.~Goldberg, and U.~Hengartner.
\newblock Louis, lester and pierre: three protocols for location privacy.
\newblock In {\em Proceedings of the 7th international conference on Privacy
  enhancing technologies}, PET'07, pages 62--76, Berlin, Heidelberg, 2007.
  Springer-Verlag.

\bibitem{PEW_Location}
K.~Zickuhr.
\newblock Three-quarters of smartphone owners use location-based services.
\newblock
  \url{http://pewinternet.org/},
  2012.

\end{thebibliography}

\end{document}